\documentclass[12pt, a4paper]{article}
\usepackage{epsf}
\usepackage{cite}
\usepackage{amsmath,amssymb}
\input{colordvi.tex}
\usepackage[usenames,dvipsnames]{color}
\usepackage[dvips]{graphicx}
\usepackage{comment}
\begin{document}

\newcommand{\lsim}{\stackrel{<}{_\sim}}
\newcommand{\gsim}{\stackrel{>}{_\sim}}
\newcommand{\rem}[1]{{ {\color{red} [[$\spadesuit$ \bf #1 $\spadesuit$]]} }}
\renewcommand{\thefootnote}{\fnsymbol{footnote}}
\setcounter{footnote}{0}


\def\thefootnote{\fnsymbol{footnote}}
\def\a{\alpha}
\def\b{\beta}
\def\c{\varepsilon}
\def\d{\delta}
\def\e{\epsilon}
\def\f{\phi}
\def\g{\gamma}
\def\h{\theta}
\def\k{\kappa}
\def\l{\lambda}
\def\m{\mu}
\def\n{\nu}
\def\p{\psi}
\def\q{\partial}
\def\r{\rho}
\def\s{\sigma}
\def\t{\tau}
\def\u{\upsilon}
\def\v{\varphi}
\def\w{\omega}
\def\x{\xi}
\def\y{\eta}
\def\z{\zeta}
\def\D{\Delta}
\def\G{\Gamma}
\def\H{\Theta}
\def\L{\Lambda}
\def\F{\Phi}
\def\P{\Psi}
\def\S{\Sigma}
\def\me{\mathrm e}

\def\o{\over}
\def\beq{\begin{eqnarray}}
\def\eeq{\end{eqnarray}}
\newcommand{\vev}[1]{ \left\langle {#1} \right\rangle }
\newcommand{\bra}[1]{ \langle {#1} | }
\newcommand{\ket}[1]{ | {#1} \rangle }
\newcommand{\bs}[1]{ {\boldsymbol {#1}} }
\newcommand{\mc}[1]{ {\mathcal {#1}} }
\newcommand{\mb}[1]{ {\mathbb {#1}} }
\newcommand{\EV}{ {\rm eV} }
\newcommand{\KEV}{ {\rm keV} }
\newcommand{\MEV}{ {\rm MeV} }
\newcommand{\GEV}{ {\rm GeV} }
\newcommand{\TEV}{ {\rm TeV} }
\def\diag{\mathop{\rm diag}\nolimits}
\def\Spin{\mathop{\rm Spin}}
\def\SO{\mathop{\rm SO}}
\def\O{\mathop{\rm O}}
\def\SU{\mathop{\rm SU}}
\def\U{\mathop{\rm U}}
\def\Sp{\mathop{\rm Sp}}
\def\SL{\mathop{\rm SL}}
\def\tr{\mathop{\rm tr}}
\def\sp{\;\;}

\def\IJMP{Int.~J.~Mod.~Phys. }
\def\MPL{Mod.~Phys.~Lett. }
\def\NP{Nucl.~Phys. }
\def\PL{Phys.~Lett. }
\def\PR{Phys.~Rev. }
\def\PRL{Phys.~Rev.~Lett. }
\def\PTP{Prog.~Theor.~Phys. }
\def\ZP{Z.~Phys. }

\begin{titlepage}

\begin{center}

\hfill UT-15-43\\
\hfill IPMU 15-0223\\
\hfill CTPU-15-28

\vskip .75in

{\Large \bf 
CMB Constraint on Dark Matter Annihilation \\[.5em]
after Planck 2015
}

\vskip .75in

{\large Masahiro Kawasaki$^{a,c}$, Kazunori Nakayama$^{b,c}$ and Toyokazu Sekiguchi$^d$}

\vskip 0.25in

\begin{tabular}{ll}
$^{a}$ &\!\! {\em Institute for Cosmic Ray Research,}\\
&{\em The University of Tokyo,  Kashiwa, Chiba 277-8568, Japan}\\
$^{b}$&\!\! {\em Department of Physics, Faculty of Science, }\\
& {\em The University of Tokyo,  Bunkyo-ku, Tokyo 133-0033, Japan}\\[.3em]
$^{c}$ &\!\! {\em Kavli IPMU (WPI), UTIAS,}\\
&{\em The University of Tokyo,  Kashiwa, Chiba 277-8583, Japan}\\
$^{d}$ &\!\! {\em Institute for Basic Science, Center for Theoretical Physics of the Universe,}\\
&{\em  Daejeon 34051, South Korea}
\end{tabular}

\end{center}
\vskip .5in

\begin{abstract}

We update the constraint on the dark matter annihilation cross section by using the recent measurements of the CMB anisotropy
by the Planck satellite.
We fully calculate the cascade of dark matter annihilation products and their effects on ionization, heating and excitation of the hydrogen,
hence do not rely on any assumption on the energy fractions that cause these effects.

\end{abstract}

\end{titlepage}


\renewcommand{\thepage}{\arabic{page}}
\setcounter{page}{1}
\renewcommand{\thefootnote}{\#\arabic{footnote}}
\setcounter{footnote}{0}
\baselineskip 0.58cm


\section{Introduction}

Dark matter (DM) constitutes more than 20\% of the present energy density of the universe.
Despite tremendous efforts to directly or indirectly detect DM particles, we still do not know its particle physics nature.
However, recent developments in experiments make constraints on DM properties severer,
especially for the so-called weakly-interacting massive particle (WIMP) DM.

In the WIMP DM scenario, the DM particle has a self-annihilation cross section of the order of the weak scale,
which can lead to a right amount of DM relic abundance consistent with observations.
The ``canonical'' value of the self-annihilation cross section to reproduce the observed amount of DM is
\begin{align}
	\langle \sigma v\rangle \simeq 3\times 10^{-26}\,{\rm cm^3s^{-1}}.  \label{can}
\end{align}
One of the stringent constraints on the DM annihilation cross section comes from the gamma-ray observations of dwarf spheroidal galaxies 
by the Fermi satellite~\cite{Ackermann:2015zua}.
The derived upper bound on the cross section is actually close to the canonical value \eqref{can} for the DM mass of $\sim 100$\,GeV
depending on the final state of the annihilation products.
Another constraint on the DM annihilation cross section is obtained from the big-bang nucleosynthesis (BBN)~\cite{Reno:1987qw,Frieman:1989fx,Jedamzik:2004ip,Hisano:2008ti,Hisano:2009rc,Kawasaki:2015yya},
which also gives stringent upper bound for the hadronic annihilation channel.

A robust constraint on the DM annihilation cross section is also obtained from the measurement of the cosmic microwave background (CMB) anisotropy.
DM annihilation around the recombination epoch injects extra energy that contributes to ionization of the neutral hydrogen
and also to heating of them, hence it modifies the standard recombination history~\cite{Shull:1982zz,Chen:2003gz,Padmanabhan:2005es}.
Thus the precise measurements of the CMB anisotropy have a high sensitivity to the amount of extra energy injection 
around the recombination epoch, which gives a robust constraint on the DM annihilation cross section~\cite{Zhang:2006fr,Mapelli:2006ej,Kanzaki:2008qb,Natarajan:2008pk,Belikov:2009qx,Galli:2009zc,Huetsi:2009ex,Cirelli:2009bb,Slatyer:2009yq,Kanzaki:2009hf,Valdes:2009cq,Hisano:2011dc,Finkbeiner:2011dx,Slatyer:2012yq,Lopez-Honorez:2013lcm,Diamanti:2013bia,Galli:2013dna,Madhavacheril:2013cna,Ade:2015xua,Slatyer:2015kla,Slatyer:2015jla}.
This constraint is robust in a sense that it does not suffer from astrophysical uncertainties, such as DM density profile in galaxies or clusters.

In this letter we update constraints on the DM annihilation cross section by using the newest data from the Planck satellite.
The Planck collaboration derived a constraint on the combination of $f_{\rm eff} \langle \sigma v\rangle$
with a parameter $f_{\rm eff}$ corresponding to energy fraction that is absorbed by the gas~\cite{Ade:2015xua}.
Many past works just left $f_{\rm eff}$ as a free parameter or used an approximation given in Ref.~\cite{Chen:2003gz}
in terms of the ionization fraction of the hydrogen $x_e$, which, however, is not always justified.\footnote{
	Refs.~\cite{Slatyer:2015kla,Slatyer:2015jla} extended the analysis of Planck~\cite{Ade:2015xua}
	to accurately calculate the effect of DM annihilation without such an approximation.
}
We adopt the method developed in our previous works \cite{Kanzaki:2008qb,Kanzaki:2009hf}
to calculate the cascade of DM annihilation products during/after the recombination, taking all the energy losses,
scatterings, ionizations and excitations into account and their effects on the CMB anisotropy
without relying on such an approximation.

\section{CMB Constraint}

Let us describe our method. 
We fully simulated how the background plasma at a redshift $z$ is affected for any initial injected energy $E$ at 
a higher redshift $z'$ taking account of all the relevant processes.
Technical details of our calculation are found in Refs.~\cite{Kanzaki:2008qb,Kanzaki:2009hf} and not repeated here.
Below we just briefly summarize our procedure.
The procedure is first to tabulate 
\begin{align}
	 \frac{d\chi_{i,h,e}^{(e)}(E,z',z)}{dz}  {\rm ~~and~~}  \frac{d\chi_{i,h,e}^{(\gamma)}(E,z',z)}{dz},
\end{align}
which respectively represent the fractions of injected electron (superscript $e$) and photon (superscript $\gamma$) energy $E$ at the redshift $z'$
that is compensated for ionization (subscript $i$), heating (subscript $h$) and excitation (subscript $e$) at the redshift $z (\leq z')$.
Then the ionization fraction of the hydrogen atom $(x_e)$ receives an additional contribution as
\begin{align}
	-\left[ \frac{dx_e}{dz}\right]_{\rm DM} = \sum_F \int_z \frac{dz'}{H(z')(1+z')}\frac{n_\chi^2(z')\langle \sigma v\rangle_F}{2n_H(z')}
	\frac{m_\chi}{E_{\rm Ry}} \frac{d\chi_i^{F}(m_\chi,z',z)}{dz},
\end{align}
where
\begin{align}
	\frac{d\chi_i^{F}(m_\chi,z',z)}{dz}=\int dE \frac{E}{m_\chi}\left[ 
		2\frac{dN_F^{(e)}}{dE} \frac{d\chi_i^{(e)}(E,z',z)}{dz} + \frac{dN_F^{(\gamma)}}{dE} \frac{d\chi_i^{(\gamma)}(E,z',z)}{dz} 
	\right],
\end{align}
$E_{\rm Ry}=13.6$\,eV is the Rydberg energy, $m_\chi$ the DM mass, $n_\chi$ the DM number density, $n_H$ the number density of the hydrogen and $F$ represents the final state of the DM annihilation. We consider $F=2\gamma$, $e^+e^-$, $\mu^+\mu^-$ and $W^+W^-$ in the following.
Here $dN_F^{(e,\gamma)}/dE$ is the electron/photon spectrum resulting from the cascade decay of the final state $F$.\footnote{
	The factor $2$ in front of $dN_F^{(e)}/dE$ accounts for the contribution from the positron.
}
This is calculated by the PYTHIA package~\cite{Sjostrand:2006za}.
The gas temperature $T_b$ is also modified in a similar manner as
\begin{align}
	-\left[ \frac{dT_b}{dz}\right]_{\rm DM} = \sum_F \int_z \frac{dz'}{H(z')(1+z')}\frac{n_\chi^2(z')\langle \sigma v\rangle_F}{3n_H(z')}
	m_\chi \frac{d\chi_h^{F}(m_\chi,z',z)}{dz},
\end{align}
where
\begin{align}
	\frac{d\chi_h^{F}(m_\chi,z',z)}{dz}=\int dE \frac{E}{m_\chi}\left[ 
		2\frac{dN_F^{(e)}}{dE} \frac{d\chi_h^{(e)}(E,z',z)}{dz} +\frac{dN_F^{(\gamma)}}{dE} \frac{d\chi_h^{(\gamma)}(E,z',z)}{dz} 
	\right].
\end{align}

The main effects of the increase of the ionization fraction on the CMB anisotropy are twofold.
One is suppression of the power spectrum at small angular scales due to the broadening of the last scattering surface.
The other is the enhancement of the polarization power spectra at low multipoles because of the
increased probability of the Thomson scattering.
Thus observations of both the CMB TT power spectrum and polarization spectra are useful to constrain DM annihilation cross section.

We included the contribution of these effects in the RECFAST code~\cite{Seager:1999bc}, a part of the CAMB code
to calculate the CMB anisotropy~\cite{Lewis:1999bs}.
We have modified the CosmoMC code~\cite{Lewis:2002ah} to include them and scan the DM mass and cross section
as well as other cosmological parameters to derive constraints on them.
We have varied $\langle \sigma v\rangle$ within $[10^{-27},10^{-23}]$~cm$^3$/sec
and $m_\chi$ within $[1,10^4]$~GeV ( $[80,10^4]$~GeV for $W^+ W^-$ chanel). 
Top-hat priors are imposed on $\langle \sigma v\rangle$ and $1/m_\chi$.

We adopted the recent Planck data of the CMB primary anisotropies (hereafter denoted as ``CMB").
Likelihood is computed based on the angular spectra of the TT+TE+EE correlations at high-$\ell$ ($\ell\ge30$) 
and TT+TE+EE+BB at low-$\ell$ ($\ell\le29$)~\cite{Aghanim:2015xee}.
In order to solve parameter degeneracy,
we optionally incorporate other cosmological data (collectively denoted as ``ext") 
including the CMB lens power spectrum from Planck~\cite{Ade:2015zua}, 
the baryon acoustic oscillation in galaxy correlation functions~\cite{Anderson:2013zyy}, 
the JLA compilation of type-Ia supernovae~\cite{Betoule:2014frx}, 
a measurement of Hubble constant $H_0=70.6 \pm 3.3$~km/s/Mpc from~\cite{Efstathiou:2013via}, 
and the CHFTLenS cosmic shear power spectrum \cite{Heymans:2012gg}. 

\begin{figure}[t]
\begin{center}
\begin{tabular}{cc}
\includegraphics[scale=1.3]{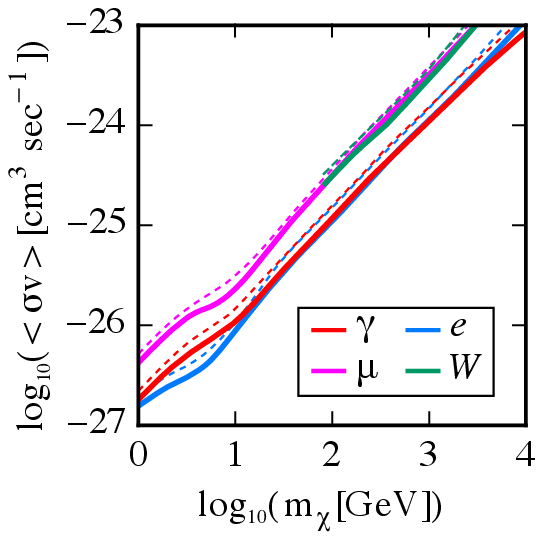} &
\includegraphics[scale=1.3]{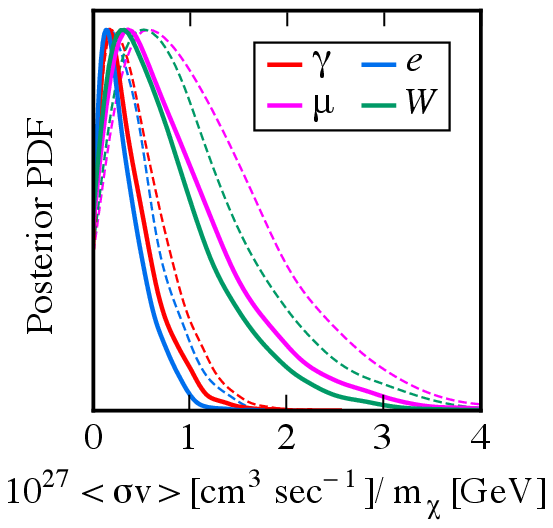}
\end{tabular}
\end{center}
\caption {
(Left) Constraints in the $\langle\sigma v\rangle$-$m_\chi$ plane.
Top left regions bounded by thick solid (thin dashed) lines are excluded 
at 95 \% from the CMB+ext (CMB-only) dataset.
Red, blue, magenta and green lines correspond to annihilation channels 
$2\gamma$, $e^+e^-$, $\mu^+\mu^-$ and $W^+W^-$, respectively. 
(Right) 1-dim posterior distributions of $\langle\sigma v\rangle/m_\chi$.
}
\label{fig:constraints}
\end{figure}

Figure~\ref{fig:constraints} plots our constraints in the $\langle\sigma v\rangle$-$m_\chi$ plane
and the 1-dim posterior distributions of a quantity $\langle\sigma v\rangle/m_\chi$ 
from the CMB-only and CMB+ext datasets. One can see from the former plot that constraints 
on WIMP annihilation in the $\langle\sigma v\rangle$-$m_\chi$ plane
virtually degenerate along constant $\langle\sigma v\rangle/m_\chi$.
As in the literature, it would be hence convenient to quote the constraints in
terms of $\langle\sigma v\rangle/m_\chi$. 
When all the data mentioned above are combined (i.e. ``CMB+ext"), 
the 95\% upper bounds on $\langle\sigma v\rangle/m_\chi$ for each annihilation channel are 
$1.3\times 10^{-27}$~cm$^3$/sec/GeV for $2\gamma$,
$1.0\times10^{-27}$~cm$^3$/sec/GeV for $e^+e^-$, 
$2.9\times10^{-27}$~cm$^3$/sec/GeV for $\mu^+\mu^-$
and $2.5\times10^{-27}$~cm$^3$/sec/GeV for $W^+W^-$.
We summarize the 95\% upper bounds on $\langle\sigma v\rangle/m_\chi$
from different datasets in Table~\ref{tab:ubounds}. 
As can be read from the table, inclusion of the ext data to CMB 
improves the constraints slightly.

\section{Discussion}

In this letter we have derived the updated CMB constraint on DM annihilation into $2\gamma$, $e^{+}e^{-}$, $\mu^{+}\mu^{-}$ and $W^{+}W^{-}$ fully taking into account the cascade of dark matter annihilation products and their effects on ionization, heating and excitation of the hydrogen.
The result can apply to various models of DM, in particular to Wino DM in supersymmetric models since Winos annihilate into $W^+ W^-$ with branching ratio almost $1$.
Thus, we can exclude the Wino DM lighter than $\sim 250$\,GeV, assuming that Wino is a dominant component of DM.\footnote{
	The thermal relic abundance of Wino can explain the present DM abundance only for the Wino mass of $\sim 3$\,TeV~\cite{Hisano:2006nn}.
	For lighter Wino, some nonthermal production mechanism, such as the decay of gravitino, is required to explain the observed DM abundance.
}

Compared with previous studies, our constraints from CMB alone are less tight than the results of 
Planck~\cite{Ade:2015xua} for the channel $e^+e^-$, in which it is assumed that energy from DM annihilation is instantaneously
converted into the gas with a constant efficiency $f_{\rm eff} \simeq 0.67$.
Admitting less model-independence, our analysis has an advantage over previous ones which assumed constant efficiency
in the point that we consistently take into account the time-evolution of the energy conversion from DM into the gas.
On the other hand, our constraints from CMB alone are largely consistent with Ref.~\cite{Slatyer:2015jla}, which does not assume a constant efficiency 
and calculated the time evolution of DM annihilation products and their effects on the gas carefully based on the methods developed in Refs.~\cite{Slatyer:2009yq,Galli:2013dna,Slatyer:2015kla}.
Thus our result may also be regarded as an independent cross-check of their method.

\begin{table}
\begin{center}
\begin{tabular}{l ||r|r|r|r}
& $2\gamma$ &  $e^+e^-$ & $\mu^+\mu^-$  & $W^+W^-$  \\
\hline
CMB-only& 1.5 & 1.4 & 3.6 & 3.2\\
CMB+ext & 1.3 & 1.0 & 2.9 & 2.5
\end{tabular}

\caption {95\% upper limits on $\langle \sigma v\rangle/m_\chi$ for 
each annihilation channel in units of $10^{-27}$~cm$^3$/sec/GeV.
}
\label{tab:ubounds}
\end{center}
\end{table}

Finally let us mention comparison with other constraints on DM annihilation.
The derived CMB constraint on DM annihilation cross section is much tighter than that from BBN for the leptonic annihilation channel.
For the hadronic annihilation channel, our result is comparable to the BBN constraint~\cite{Kawasaki:2015yya}.
The gamma-ray observation of the dwarf spheroidal galaxies~\cite{Ackermann:2015zua} also gives 
stronger constraint for the hadronic channel and comparable upper bound for the leptonic annihilation channel.
We stress that the CMB constraint is robust and does not suffer from astrophysical uncertainties compared with BBN/gamma-ray constraints.

\section*{Acknowledgments}

This work was supported by the Grant-in-Aid for Scientific Research on Scientific Research A (No.26247042 [KN]), Scientific Research C (No.25400248 [MK]), 
Young Scientists B (No.26800121 [KN]), Innovative Areas (No.26104009 [KN], No.15H05888 [KN], No.15H05889 [MK]) and World Premier International Research Center Initiative (WPI Initiative), MEXT, Japan.
TS is supported by IBS under the project code, IBS-R018-D1.
This work was supported by World Premier International Research Center Initiative (WPI Initiative), MEXT, Japan.



\end{document}